\documentclass[aps,prd,preprint,showpacs,preprintnumbers,nofootinbib,superscriptaddress]{revtex4}
\usepackage{ae}
\usepackage{bm} 
\usepackage{color}
\usepackage{amsmath}
\usepackage{amssymb}
\usepackage{amsfonts}
\usepackage{graphicx}
\usepackage{slashed}
\usepackage{wasysym}
\usepackage{setspace}

\newcommand{\Eqref}[1]{Eq.~\eqref{#1}}

\allowdisplaybreaks

\begin{document}

\setlength{\unitlength}{1mm}
\title{Laser photon merging in an electromagnetic field inhomogeneity}
\author{Holger Gies}
\author{Felix Karbstein}
\affiliation{Helmholtz-Institut Jena, Fr\"obelstieg 3, 07743 Jena, Germany}
\affiliation{Theoretisch-Physikalisches Institut, Abbe Center of Photonics, \\ Friedrich-Schiller-Universit\"at Jena, Max-Wien-Platz 1, 07743 Jena, Germany}
\author{Rashid Shaisultanov}
\affiliation{Nazarbayev University, NURIS block 9, 53 Kabanbay Batyr Ave., Astana, 010000, Republic of Kazakhstan}

\date{\today}

\begin{abstract}
We study the effect of laser photon merging, or equivalently high
harmonic generation, in the quantum vacuum subject to inhomogeneous
electromagnetic fields. Such a process is facilitated by the
effective nonlinear couplings arising from charged
particle-antiparticle fluctuations in the quantum vacuum subject to
strong electromagnetic fields. We derive explicit results for
general kinematic and polarization configurations involving
optical photons. Concentrating on merged photons in reflected
channels which are preferable in experiments for reasons of noise
suppression, we demonstrate that photon merging is typically
dominated by the competing nonlinear process of quantum reflection,
though appropriate polarization and signal filtering could
specifically search for the merging process.
As a byproduct, we devise a novel systematic expansion of the
photon polarization tensor in plane wave fields.
\end{abstract}


\maketitle

\section{Introduction} \label{sec:intro}

The discovery of quantum vacuum nonlinearities
\cite{Heisenberg:1935qt,Weisskopf,Schwinger:1951nm} under controlled
laboratory conditions using real photons or macroscopic
electromagnetic fields is a major goal of contemporary strong-field
physics. Many proposals rely on a pump-probe scheme, where a
well-controlled, say optical, photon beam probes a region of space
that is exposed to a strong field (``pump''). A typical example is given
by schemes intended to verify vacuum birefringence
\cite{Toll:1952,Baier,BialynickaBirula:1970vy,Adler:1971wn} that can
be searched for using macroscopic magnetic fields
\cite{Cantatore:2008zz,Berceau:2011zz} or with the aid of
high-intensity lasers \cite{Heinzl:2006xc}, see e.g.,
\cite{Dittrich:2000zu,Marklund:2008gj,Dunne:2008kc,DiPiazza:2011tq} for reviews.

\begin{figure}
\includegraphics[width=0.57\textwidth]{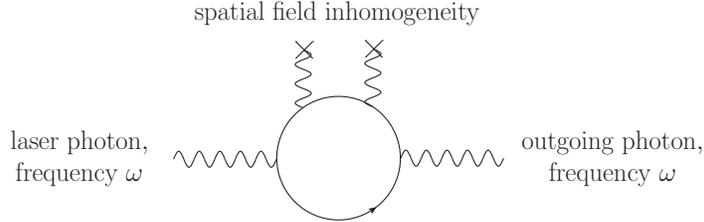} 
\caption{Typical Feynman diagram contributing to the effect of quantum reflection \cite{Gies:2013yxa}. For field strengths of the inhomogeneity well below the critical field strength (cf. main text), the leading contribution arises from diagrams with two couplings to the field inhomogeneity. As there is no energy transfer from static fields, the frequencies of the incident and outgoing photons match.}
\label{fig:QRef}
\end{figure}

As these setups require techniques such as high-purity ellipsometry
\cite{Cantatore:2008zz,Marx:2011} to separate the (small) signal from
a typically huge background, a recent proposal has focused on a
quantum-reflection scheme that facilitates a built-in noise
suppression \cite{Gies:2013yxa}. In this scheme, incident
probe photons propagate towards a spatially localized field
inhomogeneity (``pump''), as, e.g., generated in the focal spots of a
high-intensity laser system.  Even though the inhomogeneity acts similar to
an attractive potential, probe photons can be scattered backwards due to
quantum reflection.  Looking for reflected photons in the field free
region, this scenario inherently allows for a clear geometric
separation between signal and background.  First estimates of the
number of reflected photons attainable in present and near future
laser facilities look promising. Figure~\ref{fig:QRef} depicts a
typical Feynman diagram contributing to the effect.

As quantum reflection crucially relies on the presence of an
inhomogeneous pump field, it belongs to a general class of
quantum-induced interference effects
\cite{King:2013am,Tommasini:2010fb,Hatsagortsyan:2011} with the
particular property of optimizing the signal-to-noise ratio.

The pump-probe scheme is typically also reflected by the theoretical
description, in which the nonlinearities are kept for the pump-probe
interaction, but the equations are linearized with respect to the
probe propagation. In the present work, we rely again on an optical
pump-probe setup which however requires a nonlinear treatment of the
probe-field. The idea is to look for laser photon merging in the
presence of an electromagnetic field inhomogeneity.  This effect
resembles the standard nonlinear optical process of {\it second
  harmonic generation} (SHG) -- or in general {\it high harmonic
  generation} -- with the nonlinear crystal replaced by the quantum
vacuum subject to strong electromagnetic fields. Higher harmonic
generation in an electromagnetized vacuum has been discussed on the
level of the Heisenberg-Euler action in
\cite{BialynickaBirula:1981,Kaplan:2000aa,Valluri:2003sp,Fedotov:2006ii},
see also the discussion in \cite{Marklund:2006my}, or using the
constant-field polarization tensor in \cite{DiPiazza:2005jc}. Laser
photon merging in proton-laser collisions have been investigated in
detail in \cite{DiPiazza:2007cu,DiPiazza:2009cq}, where a promising
scenario has been proposed for a discovery of the merging phenomenon
that involves a nowadays conventional optical high-intensity laser at
a high-energy proton collider. A related effect is called four-wave
mixing for which also a concrete experimental proposal has been
explored in \cite{Lundin:2006aa,King:2012aw}.  The same underlying
quantum vacuum nonlinearity could even be used to radiate photons from
the focal spot of a single focused laser beam (``vacuum emission'') as
proposed in \cite{Monden:2011}. More generally, frequency mixing induced
by quantum vacuum effects has even been suggested as a sensitive probe
to search for new hypothetical particles \cite{Dobrich:2010hi}.

In the present work, we concentrate on an ``all-optical'' parameter
regime realizable with high-intensity lasers. As the signal is
expected to be very small, we again consider specifically the
kinematics of the reflection process for an appropriate
signal-to-noise reduction.  As in \cite{Gies:2013yxa}, we limit
ourselves to the study of time-independent field inhomogeneities, such
that there is no energy transfer from the field inhomogeneity. Depending on the spatial field
inhomogeneity, the propagation direction of the merged photons can
differ from that of the incident probe photons.  For the specific
reflecting kinematic situation, the merged photons can even
propagate -- somewhat counter-intuitively -- into the backward
direction.  For a straightforward comparison of the signals resulting
from quantum reflection \cite{Gies:2013yxa} and the photon merging
scenario of this work, we focus on a one-dimensional magnetic field
inhomogeneity. As is shown by an explicit calculation below, our
findings confirm the expectation that the merging process for the
reflective scenario is dominated by the quantum reflection process for
the all-optical parameter regime. Nevertheless, due to a different
polarization and frequency dependence, filtering techniques might
allow for a discovery of the merging process in this set up as well.

\begin{figure}
\includegraphics[width=0.7\textwidth]{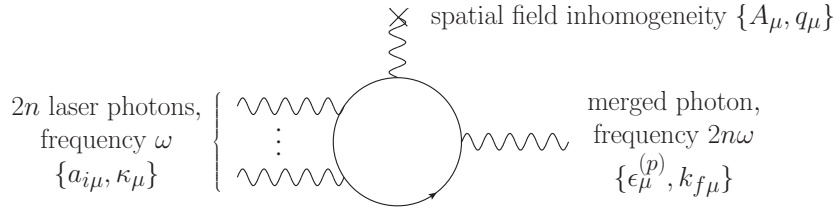} 
\caption{Cartoon of the photon merging process. In the presence of a stationary but
  spatially inhomogeneous electromagnetic field $2n$ laser photons of frequency
  $\omega$ can merge into a single photon of frequency $2n\omega$.
  Depending on the spatial field inhomogeneity, the propagation
  direction of the merged photons can differ from that of the incident
  probe photons. In curly braces we introduce our notation for the corresponding fields/polarizations and four-momenta; cf. also Eqs.~\eqref{eq:background}, \eqref{eq:Ak} and \eqref{eq:M4}, as well as Fig.~\ref{fig:perspective}.}
\label{fig:merging_cartoon}
\end{figure}

Let us briefly outline the theoretical framework of our study,
tailored to an all-optical scenario.
Optical lasers operate at frequencies $\omega\sim{\cal O}({\rm eV})$
much smaller than the electron mass $m\approx511\,{\rm keV}$,
constituting a typical scale associated with quantum effects in
quantum electrodynamics (QED), such that $\frac{\omega}{m}\ll1$.
Moreover, the maximum field strengths attainable with present and near
future laser facilities are small in comparison to the {\it critical
  field strength} $E_{\rm cr}\equiv\frac{m^2}{e}$
\cite{Heisenberg:1935qt}, i.e., $\{\frac{e{\mathfrak
    E}}{m^2},\frac{eB}{m^2}\}\ll1$, with $\mathfrak{E}$ denoting the
electric field strength of the probe laser and $B$ the peak magnetic field
strength of the spatially localized inhomogeneity.
Hence, for a given number $2n$, $n\in\mathbb{N}$, of probe laser
photons of frequency $\omega$ (wavelength
$\lambda=\frac{2\pi}{\omega}$), the dominant merging process into a
single photon of frequency $2n\omega$ is expected to arise from an
interaction of the type depicted in Fig.~\ref{fig:merging_cartoon},
exhibiting a single coupling to the (magnetic) field
inhomogeneity. Higher order couplings to the field inhomogeneity are
strongly suppressed due to the fact that $\frac{eB}{m^2}\ll1$.
Furry's theorem (charge conjugation symmetry of QED) dictates the
interaction to vanish for any odd number of couplings to the
electron-positron loop, which justifies that we have 
tailored the merging process to $2n$ laser photons. 
The dominant contribution in the weak-field limit is expected to arise from
the merging of two laser photons, described by Feynman diagrams with
four legs (cf. Fig.~\ref{fig:merging_cartoon}).

\begin{figure}[h]
\center
\includegraphics[width=0.8\textwidth]{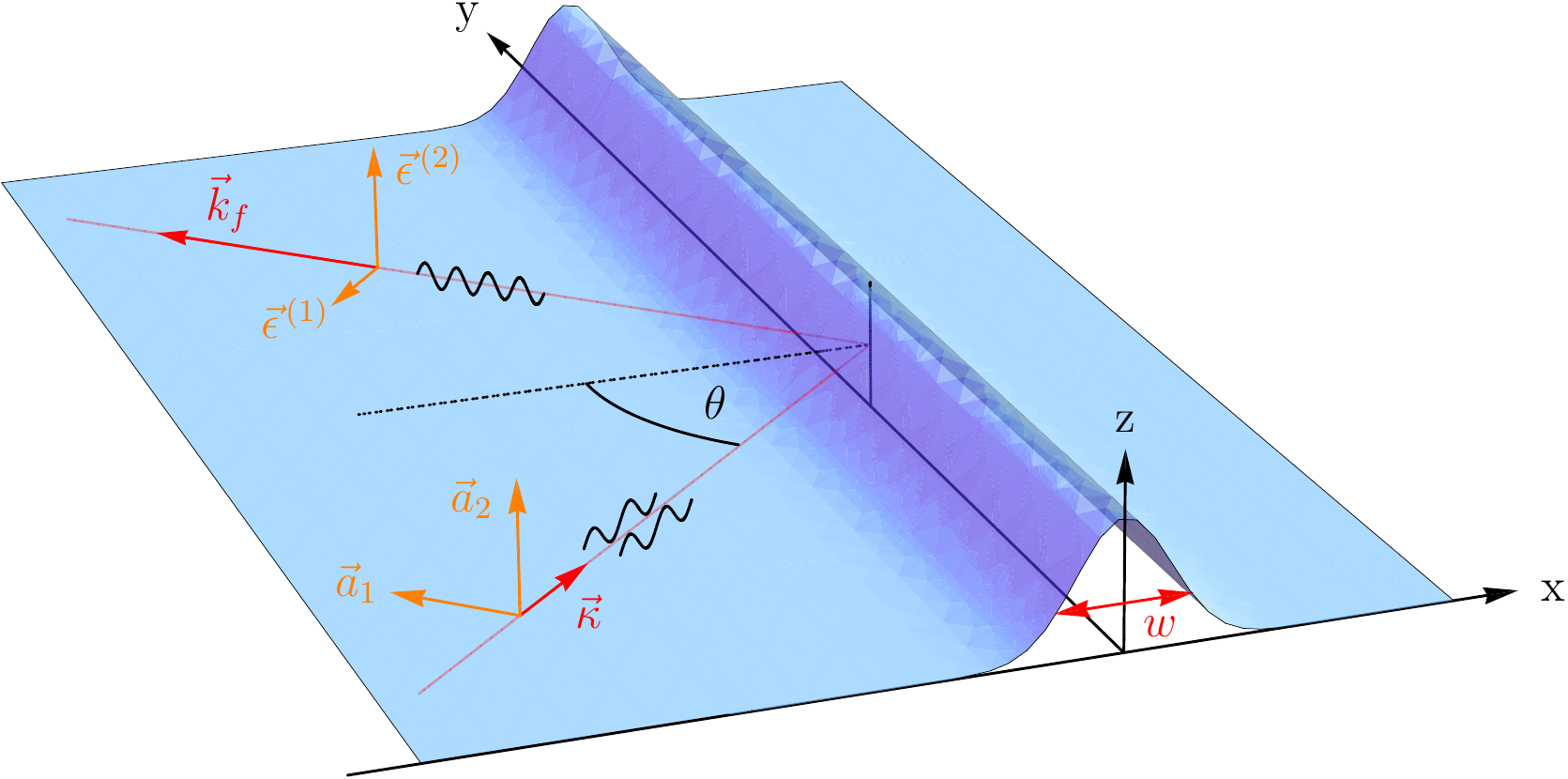} 
\caption{Schematic depiction of the two-photon merging process. Incident probe photons (wave vector $\vec{\kappa}$, energy $\kappa^0=|\vec{\kappa}|=\omega$) hit a one-dimensional field inhomogeneity $\vec{B}({\rm x})=B({\rm x})\vec{e}_{\rm z}$ of width $w$ under an angle of $\theta$. Due to nonlinear effective couplings between electromagnetic fields mediated by virtual charged particle fluctuations, the field inhomogeneity can impact incident probe photons to merge and form an outgoing photon (wave-vector $\vec{k}_f$) of twice the energy of the incident probe photons, i.e., $k_f^0=|\vec{k}_f|=2\omega$. 
Most notably, the inhomogeneity can affect the outgoing merged photons to reverse their momentum component along $\vec{e}_{\rm x}$ with respect to the incident probe photons.
The vectors $\vec{a}_1$, $\vec{a}_2$ and $\vec{\epsilon}^{\,(1)}$, $\vec{\epsilon}^{\,(2)}$ span the polarization degrees of freedom of the incident and outgoing photons, respectively. For the depiction we specialized to $\varphi=\varphi'=0$ (cf. the main text).}
\label{fig:perspective}
\end{figure}

A sketch of the geometry of the reflective scenario
of the merging process
to be investigated in this paper can be found in  Fig.~\ref{fig:perspective}.
Here we already summarize the notation to be introduced and discussed below.

The leading quantum reflection process in the perturbative regime also
arises from four leg diagrams (cf. Fig.~\ref{fig:QRef}).  While
quantum reflection necessitates at least two couplings to the field
inhomogeneity, photon merging just needs a single coupling to the
inhomogeneity. Conversely, quantum reflection can be considered as a
two-photon (one incident, one outgoing) process, whereas photon
merging involves at least three photons (two incident, one
outgoing). From this observation, it can already be anticipated
that the dependence of the observables on the various parameters
will differ between the two processes.

Notably, the merging process in Fig.~\ref{fig:merging_cartoon} can be
evaluated straightforwardly, owing to the fact that the photon
polarization tensor is explicitly known for generic monochromatic
plane wave backgrounds \cite{Baier:1975ff,Becker:1974en}.
Interpreting the plane wave background in terms of incident probe
photons of frequency $\omega$, the two open legs of the polarization
tensor can be identified with the field inhomogeneity and the outgoing
merged photon, respectively.  The polarization of the incident photons
can be controlled by adjusting the polarization of the monochromatic
plane wave background.

Our paper is organized as follows: In Sec.~\ref{sec:calculation} we
explain in detail the various steps needed to evaluate the photon
merging process.  A crucial technical step is to find a
controlled approximation to the photon polarization tensor in a plane
wave background, facilitating an analytical treatment of the photon
merging process.  Such an approximation, especially suited to the
parameters of an all-optical experimental scenario, is derived in
Sec.~\ref{subsec:Pi}.  Section~\ref{seq:Ex+Res} is devoted to the
discussion of explicit examples and results. It contains a thorough
comparison of the effects of laser photon merging and quantum reflection. 
We end with conclusions and an outlook in Sec.~\ref{seq:Con+Out}.

\section{Calculation} \label{sec:calculation}

\subsection{Photon polarization tensor in plane wave field} \label{subsec:Pi}

We briefly recall and summarize the basic structure of the photon
polarization tensor in a generic, elliptically polarized monochromatic
plane wave background \cite{Baier:1975ff,Becker:1974en}. The latter is parametrized by the following gauge potential in Coulomb-Weyl gauge
\begin{equation}
{\cal A}_\mu(x)={\mathfrak a}_{1}a_{1\mu}\cos(\kappa x)+{\mathfrak a}_{2}a_{2\mu}\sin(\kappa x), \label{eq:background}
\end{equation}
with $\mathcal{A}_0=0$, $\kappa^2=0$ and
$a_1\kappa=a_2\kappa=a_1a_2=0$. Moreover, we will use
the frequency $\omega\equiv\kappa^0$.  The four-vectors $a_{i\mu}$ with
$i\in\{1,2\}$ are normalized to unity, i.e., $a_i^2=1$, and the field
amplitude is encoded in the coefficients ${\mathfrak a}_i\geq0$.  Our
metric convention is $g_{\mu \nu}=\mathrm{diag}(-1,+1,+1,+1)$, and we use $c=\hbar=1$.

For the normalized plane wave field strength in momentum space,
we introduce $f^{\mu\nu}_i=\kappa^\mu a_i^\nu-\kappa^\nu a_i^\mu$. 
In the following, we will frequently use the shorthand notation $(kf_i)^\mu=k_\nu f^{\nu\mu}_i$. 

In momentum space the photon polarization tensor mediates between two
four-momenta $k_1$ and $k_2$.  Since the wave~\eqref{eq:background} is
characterized by the single four-momentum $\kappa$ and a change in the
incident momentum is determined by an interaction with the wave, the
kinematics are such that $k_2=k_1+C\kappa$, with scalar constant $C$
\cite{Baier:1975ff}.  Correspondingly, $\kappa k_2=\kappa
k_1\equiv\kappa k$ and also $(k_1f_i)^\mu=(k_2f_i)^\mu=(kf_i)^\mu$.

Following \cite{Baier:1975ff}, the associated photon
polarization tensor can then be compactly represented as
\begin{equation}
 \Pi^{\mu\nu}(k_1,k_2)=c_1\Lambda_1^\mu\Lambda_2^\nu+ c_2\Lambda_2^\mu\Lambda_1^\nu+c_3\Lambda_1^\mu\Lambda_1^\nu+c_4\Lambda_2^\mu\Lambda_2^\nu+c_5\Lambda_3^\mu\Lambda_4^\nu, \label{eq:PIstructure}
\end{equation}
with scalar coefficients $c_j(k_1,k_2)$, $j\in\{1,\ldots,5\}$.
The tensor structure is encoded in products of the normalized four vectors
\begin{multline}
 \Lambda_i^\mu=\frac{(kf_i)^\mu}{(\kappa k)}=a_i^\mu-\frac{(ka_i)}{(\kappa k)}\kappa^\mu \quad \text{for} \quad i\in\{1,2\},\\
 \Lambda_3^\mu=\frac{\kappa^\mu k_1^2-k_1^\mu(\kappa k)}{(\kappa k)\sqrt{-k_1^2}}, \quad \Lambda_4^\mu=\frac{\kappa^\mu k_2^2-k_2^\mu(\kappa k)}{(\kappa k)\sqrt{-k_2^2}}, \label{eq:Lambdas}
\end{multline}
fulfilling
$\Lambda_1^2=\Lambda_2^2=\Lambda_3^2=\Lambda_4^2=1$. This tensor
structure guarantees that $\Pi^{\mu\nu} (k_1,k_2)$ satisfies the
Ward identities $k_{1,\mu}\Pi^{\mu\nu} (k_1,k_2)=\Pi^{\mu\nu}(k_1,k_2)k_{2,\nu}=0$.

Apart from a trivial overall factor of $\alpha=e^2/(4\pi)$,
the coefficients $c_j$ depend on the kinematic variables $k_1$, $k_2$
and $\kappa$ as well as the electron mass $m$, and account for
the entire field strength dependence. The latter dependence is most
conveniently expressed in terms of the two invariant intensity
parameters $\xi_i=\frac{e\mathfrak{a}_i}{m}$ with $i\in\{1,2\}$.
In Coulomb-Weyl gauge, the amplitude $\mathfrak{a}_i$ is intimately related to the amplitude
of the associated electric field $\mathfrak{E}_i$ via
$\mathfrak{a}_i=\frac{\mathfrak{E}_i}{\omega}$, such that -- in terms
of parameters directly accessible in the lab -- we have
$\xi_i=\frac{e\mathfrak{E}_i}{m\omega}$.

In consequence of Furry's theorem, the field dependence can be encoded in
$\xi_1^2$, $\xi_2^2$ and $\xi_1\xi_2$, i.e., combinations even in the
charge $e$, only.  It is moreover helpful to introduce the
dimensionless parameter $\lambda=-\frac{\kappa k}{2m^2}$,
parametrizing the relative momenta of the involved photons. 
In summary, the relevant dimensionless parameters for the
off-shell  polarization tensor in a plane wave field are given by
\begin{equation}
\xi_i=\frac{e\mathfrak{E}_i}{m\omega}, \quad \lambda=-\frac{\kappa k}{2m^2}, \quad \frac{k_1 k_2}{4m^2}, 
\end{equation}
where the last parameter characterizes the relative momenta of the in- and outgoing photon legs.

In the following, we are only interested in a situation with actual interactions
with the plane wave field~\eqref{eq:background} and thus omit the zero
field contribution in \Eqref{eq:PIstructure}.\footnote{More precisely,
  the coefficients $c_j$ provided in the following correspond to the
  quantity $\Pi^{\mu\nu}(A)-\Pi^{\mu\nu}(A=0)$.  The zero field term
  can be included straightforwardly, noting that
  $g^{\mu\nu}-\frac{k_1^\mu
    k_1^\nu}{k_1^2}=\Lambda_1^\mu\Lambda_1^\nu+\Lambda_2^\mu\Lambda_2^\nu+\Lambda_3^\mu\Lambda_3^\nu$. \label{ftnt}}

The coefficients $c_j$ generically decompose into an {\it elastic}
part characterized by zero momentum exchange with the wave and an {\it
  inelastic} part with finite momentum exchange.  The latter part is
made up of an infinite number $l\in\mathbb{Z}\setminus\{0\}$ of
contributions with momentum transfer $2\kappa l$ to be associated with
the absorption/release of $2l$ laser photons.  Correspondingly, we
write
\begin{equation}
 c_j=i(2\pi)^4m^2\frac{\alpha}{\pi}\Bigl[\delta(k_1-k_2)G_j^0+\sum_{l\in\mathbb{Z}\setminus\{0\}}\delta(k_1-k_2-2l\kappa)G_j^l\Bigr], \label{eq:c_j}
\end{equation}
where the dimensionless coefficients $G_j^l(k_1,k_2)$, with
$l\in\mathbb{Z}$, are most conveniently represented in terms of double
parameter integrals that cannot be tackled analytically in a
straightforward way.  One of the integrals is over a proper-time type
parameter $\rho\in [0,\infty[$, and the other one over an additional parameter $\nu\in [-1,1]$
related to the momentum routing in the loop.

In order to state them most compactly, it is convenient to define
\begin{multline}
 A=\frac{1}{2}\Bigl(1-\frac{\sin^2\rho}{\rho^2}\Bigr),\quad A_0=\frac{1}{2}\rho(\partial_\rho A),\quad A_1=A+2A_0, \\
 z=\frac{2(\xi_1^2-\xi_2^2)}{|\lambda|(1-\nu^2)}\rho A_0,\quad  y=\frac{2(\xi_1^2+\xi_2^2)}{|\lambda|(1-\nu^2)}\rho A. \label{eq:defs}
\end{multline}
Taking these definitions into account, the explicit expressions for $G_j^l$ read
\begin{equation}
 G_j^l=\int_{-1}^1{\rm d}\nu\int_0^\infty\frac{{\rm d}\rho}{\rho}\,{\rm e}^{-i\phi_0\rho}\,g_j^l\,{\rm e}^{-iy}, \label{eq:Gs}
\end{equation}
where
\begin{equation}
 \phi_0=\frac{2}{|\lambda|(1-\nu^2)}\Bigl[1-i\epsilon+\frac{k_1k_2}{4m^2}(1-\nu^2)\Bigr],
\end{equation}
with $\epsilon\to0^+$, and
\begin{align}
 g_1^l&=\xi_1\xi_2\Bigl(2\,{\rm sign}(\lambda)\frac{1+\nu^2}{1-\nu^2}\,\rho A_0-A_1\frac{l}{z}\Bigr)i^lJ_l(z), \nonumber\\
 g_2^l&=g_1^l\left(A_0\to-A_0,z\to z,A_1\to A_1\right), \nonumber\\
 g_3^l&=\Bigl(\xi_1^2A_1-\frac{\xi_1^2-\xi_2^2}{1-\nu^2}\sin^2\rho\Bigr)i^{l}\bigl(J_l(z)-iJ_l'(z)\bigr)+\xi_1^2\frac{1+\nu^2}{1-\nu^2}\sin^2\rho\, i^lJ_l(z) \nonumber\\
      &\quad +\frac{1}{4}\Bigl(\frac{k_1k_2}{m^2}-\frac{i|\lambda|(1-\nu^2)}{\rho}\Bigr)i^l\bigl(J_l(z)-\delta_{l0}\,{\rm e}^{iy} \bigr), \nonumber\\
 g_4^l&=g_3^l\left(\xi_1^2\leftrightarrow\xi_2^2\right)(-1)^l, \nonumber\\
 g_5^l&=-\frac{\sqrt{k_1^2k_2^2}}{4m^2}(1-\nu^2)i^l\bigl(J_l(z)-\delta_{l0}\,{\rm e}^{iy} \bigr), \label{eq:gs}
\end{align}
for $l\in\mathbb{Z}$. Here, $J_l(z)$ denotes the Bessel function of
the first kind, and $\delta_{ll'}$ is the Kronecker delta.
Equations~\eqref{eq:PIstructure}-\eqref{eq:gs} constitute the full
expression of the photon polarization tensor in a generic plane wave
background of type~\eqref{eq:background}
\cite{Baier:1975ff,Becker:1974en}; see \cite{Meuren:2013oya} for
a more recent derivation and an alternative representation.
Noteworthily, whenever one of the momenta $k_1$ and $k_2$ is on the
light cone, i.e., either $k_1^2=0$ or $k_2^2=0$, the coefficients
$G_5^l$ vanish for all $l\in\mathbb{Z}$, such that $c_5=0$. Except for
the zero field contribution (cf. footnote~\ref{ftnt}), the tensor
structure of the photon polarization tensor {under these conditions
  can be written entirely in terms of the four-vectors $\Lambda_1^\mu$
  and $\Lambda_2^\mu$.

For completeness, note that for a circularly polarized plane wave
background, corresponding to the choice of $\xi_1=\xi_2$, we have
$z=0$.  Hence, taking into account that $J_l(z)\sim z^{|l|}$
[cf. \Eqref{eq:Jseries} below], the only nonvanishing
contributions~\eqref{eq:gs} are those with $l\in\{0,\pm1\}$,
corresponding to the possibility of an elastic interaction and an
interaction involving the emission/absorption of just two photons from
the circularly polarized wave.  The physical reason for this is that a
circularly polarized wave has definite chirality, such that
transitions are only possible without a change in the chirality of the
incident photon $(l=0)$ or with a reversal of its chirality ($l=\pm1$)
\cite{Baier:1975ff,Becker:1974en}.

As the expressions are rather cumbersome, we subsequently aim at an
approximation particularly suited for all-optical experiments.  Our
strategy to achieve this relies on series expansions of the expression
$g_j^l\,{\rm e}^{-iy}$ in the integrand of \Eqref{eq:Gs}, such that
both integrals can be performed explicitly and handy approximations
for the polarization tensor are obtained. Similar expansion
strategies have recently also led to new analytical insights into
the well-known polarization tensor for constant fields \cite{Karbstein:2013ufa}.

For this purpose it is particularly helpful to note that $A$ and $A_0$
have the following infinite series representations
[cf. \Eqref{eq:defs}],
\begin{equation}
 A=\frac{\rho^2}{6}\sum_{n=0}^\infty A^{(2n)}\rho^{2n},\quad  A_0=\frac{\rho^2}{6}\sum_{n=0}^\infty (1+n)A^{(2n)}\rho^{2n}, \label{eq:As_series}
\end{equation}
with $A^{(2n)}=\frac{3}{2}\frac{(2i)^{2n+4}}{(2n+4)!}$; our definitions are such that $A^{(0)}=1$.

The above series representations suggest to define
\begin{equation}
 \zeta^\pm\equiv\frac{(\xi_1^2\pm\xi_2^2)\rho^3}{3|\lambda|(1-\nu^2)}
\end{equation}
and to rewrite the quantities $y$ and $z$ as follows
\begin{equation}
  y=\zeta^+\sum_{n=0}^\infty A^{(2n)}\rho^{2n},\quad z=\zeta^-\sum_{n=0}^\infty (1+n)A^{(2n)}\rho^{2n}. \label{eq:yzseries}
\end{equation}

Another important ingredient in our approach is the series
representation of $J_l(z)$, which, for $l\in\mathbb{Z}$, reads
(cf. formulae 8.404 and 8.440 of \cite{Gradshteyn})
\begin{equation}
 J_{l}(z)=\sum_{j=0}^{\infty}\frac{(-1)^j[{\rm sign}(l)]^l}{j!\,(|l|+j)!}\left(\frac{z}{2}\right)^{|l|+2j}\quad {\rm for}\quad |{\rm arg}(z)|<\pi\,, \label{eq:Jseries}
\end{equation}
where $[{\rm sign}(l)]^l=1$ for $l=0$ is implicitly understood.
Inserting \Eqref{eq:yzseries} into \Eqref{eq:Jseries}, all the Bessel
functions occurring in \Eqref{eq:gs} can be expanded in powers of $\zeta^-$ and $\rho^2$.
Analogously, factors of ${\rm e}^{-iy}$ can be expanded in powers of
$\zeta^+$ and $\rho^2$.

In the following, let us
assume that $|\frac{k_1k_2}{4m^2}|<1$, which is well compatible with
an all-optical experimental scenario.  Building on this assumption,
and resorting to the identity $\int_0^\infty \frac{{\rm
    d}\rho}{\rho}\,{\rm
  e}^{-i\phi_0\rho}\,\rho^{l+1}=l!\bigl(\frac{-i}{\phi_0}\bigr)^{l+1}$
for $l\in\mathbb{N}_0$, we obtain
\begin{equation}
 \int_0^\infty \frac{{\rm d}\rho}{\rho}\,{\rm e}^{-i\phi_0\rho}\,\rho^{l+1}=l!\left(-\frac{i}{2}|\lambda|(1-\nu^2)\right)^{l+1}\sum_{n=0}^\infty\binom{n+l}{n}\left(-\frac{k_1k_2}{4m^2}(1-\nu^2)\right)^n  .
\end{equation}
Having implemented the above expansions, the polarization tensor can formally be written in terms of multiple infinite sums.
Noteworthily, all $\nu$ integrals are of the following type
\begin{align}
 \int_{-1}^1{\rm d}\nu\,(1-\nu^2)^n&=\frac{2^{2n+1}(n!)^2}{(2n+1)!}, \nonumber\\
 \int_{-1}^1{\rm d}\nu\,(1+\nu^2)(1-\nu^2)^n&=\left(1+\frac{1}{2n+3}\right)\int_{-1}^1{\rm d}\nu\,(1-\nu^2)^n,
\end{align}
with $n\in\mathbb{N}_0$, and can straightforwardly be performed explicitly for each contribution.

Thus, with the collective notation $\xi^2\in\{\xi_1^2,\xi_2^2,\xi_1\xi_2\}$ a generic
contribution to the photon polarization tensor reads
\begin{multline}
 \int_{-1}^1{\rm d}\nu\int_0^\infty \frac{{\rm d}\rho}{\rho}\,{\rm e}^{-i\phi_0\rho}\Bigl(\frac{\xi^2\rho^{2}}{6}\Bigr)^s\rho^{l}(\zeta^+)^n(\zeta^-)^j
 \left\{
 \begin{array}{c}
  1 \\
  1-\nu^2 \\
  \frac{1}{1-\nu^2}\\
  \frac{1+\nu^2}{1-\nu^2}
 \end{array}
 \right\} \\
=\Bigl(-\frac{2\xi^2\lambda^2}{3}\Bigr)^{s}\Bigl(i\frac{2(\xi_1^2+\xi_2^2)\lambda^2}{3}\Bigr)^n\Bigl(i\frac{2(\xi_1^2-\xi_2^2)\lambda^2}{3}\Bigr)^j\bigl(-2i|\lambda|\bigr)^{l}\\
\times c(n,j,s,l)
 \left\{
 \begin{array}{c}
  1 \\
  1-\frac{1}{4(n+j+s)+2l+3} \\
  1+\frac{1}{2}\frac{1}{2(n+j+s)+l}\\
  1+\frac{1}{2(n+j+s)+l}
 \end{array}
 \right\} 
\Bigl(1+{\cal O}(\tfrac{k_1k_2}{4m^2})+{\cal O}(\lambda^2)\Bigr) , \label{eq:genblock}
\end{multline}
with integers $\{l,n,j\}\in\mathbb{N}_0$ and $s\in\{0,1\}$, fulfilling
$l+n+j+s>0$. The components in the columns in braces exhaust all possible
types of occurring $\nu$ integrands. The explicit
expression for the numeric coefficient in \Eqref{eq:genblock} is
\begin{equation}
 c(n,j,s,l)=\frac{2[3(n+j)+2s+l-1]!\{[2(n+j+s)+l]!\}^2}{[4(n+j+s)+2l+1]!}.
\end{equation}
Both integrations can be carried out, and \Eqref{eq:genblock} provides
us with the full numeric prefactor for given integers $l$, $n$, $j$
and $s$ at leading order in a double expansion in
$|\frac{k_1k_2}{4m^2}|\ll1$ and $|\lambda|\ll1$, both
corresponding to a soft-photon limit.
Most importantly, the parameters} $\xi_i$ never come alone but
always appear in combination with a factor of $\lambda$. This
implies that any perturbative expansion of the photon polarization
tensor in plane wave backgrounds which is superficially in powers of
$\xi^2$ in fact amounts to an expansion in the combined parameter
$\xi^2\lambda^2$. This is of substantial practical relevance, as
optical high-intensity lasers are entering the regime $\xi\gg1$.
Still the present expansion remains valid as long as
$\xi^2\lambda^2\ll 1$ which is typically well satisfied for
contemporary optical high-intensity lasers.
First indications of a larger validity regime of the naive ``small-$\xi$'' expansion had
already been observed in \cite{DiPiazza:2009cq}. Our all-order series expansion of the
polarization tensor now clarifies the systematics of the underlying
physical parameter regimes.

Correspondingly, the photon polarization tensor can be organized in
terms of an expansion in the dimensionless quantities
$\frac{k_1k_2}{4m^2}$, $\lambda$ and $\xi^2\lambda^2$.  In particular,
the leading contributions to \Eqref{eq:c_j} are of ${\cal
  O}(\xi^2\lambda^2)$ and read
\begin{align}
 G_1^0&=-G_2^0=\frac{32}{315}\xi_1\xi_2\lambda^2i\lambda\Bigl(1+{\cal O}(\tfrac{k_1k_2}{4m^2})+{\cal O}(\lambda^{2})\Bigr), \nonumber\\
 G_3^0&=-\frac{2}{45}\left(4\xi_1^2\lambda^2+7\xi_2^2\lambda^2\right)\Bigl(1+{\cal O}(\tfrac{k_1k_2}{4m^2})+{\cal O}(\lambda^{2})\Bigr),  \nonumber\\
 G_4^0&=G_3^0\left(\xi_1^2\leftrightarrow\xi_2^2\right), \nonumber\\
 G_5^0&=-\frac{8}{105}\frac{\sqrt{k_1^2k_2^2}}{4m^2}(\xi_1^2\lambda^2+\xi_2^2\lambda^2)\Bigl(1+{\cal O}(\tfrac{k_1k_2}{4m^2})+{\cal O}(\lambda^{2})\Bigr),  \label{eq:G_j^0}\\
\intertext{and}
 G_1^{\pm 1}&=G_2^{\pm 1}=\pm\frac{i}{15}\,\xi_1\xi_2\lambda^2\Bigl(1+{\cal O}(\tfrac{k_1k_2}{4m^2})+{\cal O}(\lambda^{2})\Bigr), \nonumber\\
 G_3^{\pm 1}&=\frac{1}{45}\left(4\xi_1^2\lambda^2-7\xi_2^2\lambda^2\right)\Bigl(1+{\cal O}(\tfrac{k_1k_2}{4m^2})+{\cal O}(\lambda^{2})\Bigr),  \nonumber\\
 G_4^{\pm 1}&=-G_3^{\pm 1}\left(\xi_1^2\leftrightarrow\xi_2^2\right), \nonumber\\
 G_5^{\pm 1}&=\frac{4}{105}\frac{\sqrt{k_1^2k_2^2}}{4m^2}(\xi_1^2\lambda^2-\xi_2^2\lambda^2)\Bigl(1+{\cal O}(\tfrac{k_1k_2}{4m^2})+{\cal O}(\lambda^{2})\Bigr), \label{eq:G_j^1}
\end{align}
whereas the leading contributions to $G^l_j$ with $|l|\geq2$ scale as
$\sim(\xi^2\lambda^2)^{|l|}$ and thus are at least of ${\cal
  O}((\xi^2\lambda^2)^2)$.
Plugging these terms into Eqs.~\eqref{eq:PIstructure}-\eqref{eq:c_j},
we obtain a compact approximation to the photon polarization tensor
for a generic, elliptically polarized plane wave background in the
parameter regime where
$\{\xi^2\lambda^2,|\lambda|,|\frac{k_1k_2}{4m^2}|\}\ll1$.  The above
findings imply that the infinite sum in \Eqref{eq:c_j} at ${\cal
  O}(\xi^2\lambda^2)$ receives contributions only for $l=\pm1$.
Hence, the persistent inelastic interactions can be associated with
the absorption/release of just two laser photons.

As a particular example, we consider
the special case of an incoming on-shell photon with $k_1^\mu=\omega_1(1,\vec{k}_1/|\vec{k}_1|)$, fulfilling
$k_1^2=0$. In this case, the parameter $\lambda$ can be written as
$\lambda\to\frac{\omega\omega_1}{2m^2}\bigl(1-\cos\varangle(\vec{\kappa},\vec{k}_1)\bigr)$,
such that
\begin{equation}
 \lambda^2\xi^2 \quad\to\quad \Bigl(\frac{e\mathfrak{E}}{m^2}\Bigr)^2\frac{\omega_1^2}{4m^2}\bigl(1-\cos\varangle(\vec{\kappa},\vec{k}_1)\bigr)^2,
\end{equation}
where we employed the shorthand notation
$\mathfrak{E}^2\in\{\mathfrak{E}^2_1,\mathfrak{E}^2_2,\mathfrak{E}_1\mathfrak{E}_2\}$. Obviously,
the dependence on the frequency $\omega$ of the plane wave background
drops out and the combination $\lambda^2\xi^2$ becomes $\omega$
independent.  Correspondingly, the photon polarization tensor at ${\cal O}(\xi^2\lambda^2)$ in the
limit $\omega\to0$ is obtained straightforwardly in this case:
It is given by \Eqref{eq:PIstructure} with $c_5=0$ [see the remarks
  below \Eqref{eq:gs}], and the projectors~\eqref{eq:Lambdas} and
other coefficients~\eqref{eq:c_j} specialized to $\omega=0$.
Obviously, it only features an elastic contribution and its
coefficients [cf. \Eqref{eq:c_j}] are given by
\begin{equation}
 c_j\quad\to\quad i(2\pi)^4m^2\frac{\alpha}{\pi}\delta(k_1-k_2)\tilde G_j, \label{eq:c_jCrossed}
\end{equation}
with $\tilde G_j\equiv\bigl[G_j^0+G_j^{+1}+G_j^{-1}\bigr]\big|_{\omega=0}$ and $j\in\{1,\ldots,4\}$. 
Inserting the explicit expressions from Eqs.~\eqref{eq:G_j^0} and \eqref{eq:G_j^1} into \Eqref{eq:c_jCrossed}, we obtain $\tilde G_1=\tilde G_2=0$ as well as $\tilde G_3=-\frac{28}{45}\xi_2^2\lambda^2$ and $\tilde G_4=-\frac{16}{45}\xi_2^2\lambda^2$.
As expected the dependence on $\xi_1$ completely drops out and the polarization tensor in this limit eventually depends only on the single field strength $\mathfrak{E}_2$. Recall that the electromagnetic field components follow by differentiations of the four-vector potential~\eqref{eq:background}, which explains why the electric field $\mathfrak{E}_2$, persists even though it comes along with a factor of $\sin(\kappa x)$ in \Eqref{eq:background}.
Putting everything together, we finally obtain
\begin{equation}
 \Pi^{\mu\nu}(k_1,k_2) \ \ \to\ \ -i(2\pi)^4\delta(k_1-k_2)\frac{\alpha}{\pi}\,\omega_1^2\bigl(1-\cos\varangle(\vec{\kappa},\vec{k}_1)\bigr)^2\Bigl(\frac{e\mathfrak{E}_2}{m^2}\Bigr)^2\biggl[ \frac{7}{45}\Lambda_1^\mu\Lambda_1^\nu+\frac{4}{45}\Lambda_2^\mu\Lambda_2^\nu\biggr]. \label{eq:PIstructureCrossed}
\end{equation}
This reproduces the photon polarization tensor for constant crossed fields at ${\cal
  O}\bigl((\frac{e\mathfrak{E}}{m^2})^2\bigr)$ and on-the-light-cone
dynamics \cite{narozhnyi:1968,ritus:1972}.

\subsection{Laser photon merging} \label{subsec:photonmerging}

For a given laser photon polarization, i.e., a particular
choice of the monochromatic plane wave
background~\eqref{eq:background}, the photon merging amplitude
depends on both the explicit expression for the field inhomogeneity
and the polarization state $\epsilon_\mu^{*(p)}(k)$ of the outgoing
photon, with $p$ labeling the two transverse photon polarizations.  It
is given by \cite{Yakovlev:1967}
\begin{equation}
 {\cal M}^{(p)}(k)=\frac{\epsilon_\mu^{*(p)}(k)}{\sqrt{2k^0}}\int\frac{{\rm d}^4q}{(2\pi)^4}\,\Pi^{\mu\nu}(k,q)A_\nu(q)\,, \label{eq:M}
\end{equation}
where $A_\nu(q)=\int_{x}\,{\rm e}^{-ixq}A_\nu(x)$ is the Fourier
transform of the gauge field representing the inhomogeneous electromagnetic field 
in position space; the star symbol $^*$ denotes complex
conjugation.  The explicit expression for $k^\mu=(k^0,\vec{k})$
depends of course on the specific merging process to be
considered. For the merging of $2n$ laser photons of frequency
$\omega$ in a static field, momentum conservation and the fact that
the outgoing photon is real and propagates on the light cone imply
that $k^0=|\vec{k}|=2n\omega$.  Moreover, given this condition, the
coefficient $c_5$ in \Eqref{eq:PIstructure} vanishes [cf. below
  \Eqref{eq:gs}], such that the tensor structure of
$\Pi^{\mu\nu}(q,k)$ can be expressed solely in terms of
$\Lambda_1^\mu$ and $\Lambda_2^\mu$.

As outlined in detail above, in this article we limit ourselves to the
study of the merging process in a static magnetic field.  We consider
field inhomogeneities of the form $\vec{B}(x)=B(x)\vec{e}_B$, such
that the direction of the magnetic field $\vec{e}_B$ is fixed globally
and only its amplitude is varied.  More specifically, we set
$\vec{e}_B=\vec{e}_{\rm z}$ and focus on a one dimensional spatial
inhomogeneity in $\rm x$ direction, i.e., $B(x)\to B({\rm x})$, such
that $\vec{\nabla}B({\rm x})\sim\vec{e}_{\rm x}$.  The wave vector of
the laser photons is assumed to be $\vec{\kappa}={\kappa}_{\rm
  x}\vec{e}_{\rm x}+{\kappa}_{\rm y}\vec{e}_{\rm y}$, i.e., the
incident laser photons do not have a momentum component parallel to
the magnetic field (cf. Fig~\ref{fig:perspective}). Even if they
had, such a component would not be affected due to translational
invariance along the ${\rm z}$ direction.

Utilizing $\kappa^2=0$ it is convenient to introduce the angle parameter
$\theta\in[0\ldots\frac{\pi}{2}]$ and write
$\kappa^\mu=\omega(1,\cos\theta,\sin\theta,0)$ with $\omega>0$.
Correspondingly, the orthogonality relations
$a_1\kappa=a_2\kappa=a_1a_2=0$ imply that the parametrization of the
orthonormal vectors $a_1^\mu$ and $a_2^\mu$ just requires one
additional angle parameter which we denote by
$\varphi\in[0\ldots2\pi)$.  We write
\begin{align}
 a_1^\mu&=(0,-\sin\theta\cos\varphi,\cos\theta\cos\varphi,-\sin\varphi), \nonumber\\
 a_2^\mu&=(0,-\sin\theta\sin\varphi,\cos\theta\sin\varphi,\cos\varphi), \label{eq:a_12}
\end{align}
i.e., our conventions are such that the spatial components of
$\kappa^\mu$, $a_1^\mu$ and $a_2^\mu$ form a right-handed trihedron
(cf. Fig~\ref{fig:perspective}).  The choice of $\theta$ fixes the
propagation direction $\vec{\kappa}$ of the incident photons relative
to the inhomogeneity, while $\varphi$ controls the orientation of the
vectors $\vec a_1$ and $\vec a_2$ spanning the spatial subspace
transverse to $\vec{\kappa}$.

A convenient choice for the four-vector potential giving rise to a
magnetic field of the desired type is
\begin{equation}
 A^\mu(x)=A({\rm x})e^\mu_{\rm y}, \quad\text{with}\quad A({\rm x})=\int^{\rm x}{\rm d}{\rm x}'\,B({\rm x}'), \label{eq:Ax}
\end{equation}
where we have defined $e^\mu_{\rm y}\equiv(0,\vec{e}_{\rm y})$.
The lower limit of the integral is left unspecified as it does
not have any observable consequences and thus can be chosen
arbitrarily.  Finally, a Fourier transform of \Eqref{eq:Ax} yields the
momentum space representation of the four-vector potential as needed
in \Eqref{eq:M},
\begin{equation}
 A^\mu(q)=(2\pi)^3\delta(q_0)\delta(q_{\rm y})\delta(q_{\rm z})A(q_{\rm x})e^\mu_{\rm y}, \quad\text{with}\quad A(q_{\rm x})=\int_{-\infty}^\infty{\rm d}{\rm x}\,{\rm e}^{-i{\rm x}q_{\rm x}}\,A({\rm x}) . \label{eq:Ak}
\end{equation}

Plugging this expression into \Eqref{eq:M} and introducing $\bar q^\mu\equiv(0,q_{\rm x}\vec{e}_{\rm x})$, the photon merging amplitude can be simplified significantly and reads
\begin{equation}
 {\cal M}^{(p)}(k)=\frac{\epsilon_\mu^{*(p)}(k)}{\sqrt{2k^0}}\int\frac{{\rm d}q_{\rm x}}{2\pi}\,\Pi^{\mu2}(k,\bar q)\,A(q_{\rm x})\,. \label{eq:M1}
\end{equation}
Substituting $k_2\to\bar q$ into the expressions for $\Lambda_1^\mu$ and $\Lambda_2^\mu$ in
\Eqref{eq:Lambdas} we obtain together with \Eqref{eq:a_12}
\begin{align}
 \Lambda_1^\mu&=\bigl(\tan\theta\cos\varphi,0,\tfrac{\cos\varphi}{\cos\theta},-\sin\varphi\bigr), \nonumber\\
 \Lambda_2^\mu&=\bigl(\tan\theta\sin\varphi,0,\tfrac{\sin\varphi}{\cos\theta},\cos\varphi\bigr). \label{eq:lambda12}
\end{align}
Analogously to \Eqref{eq:c_j}, we write 
\begin{equation}
 \Pi^{\mu2}(k,\bar q)=(2\pi)^4\sum_{l\in\mathbb{Z}}\delta(k-\bar q-2l\kappa)\Pi^{\mu2}_l(k,\bar q), \label{eq:Pi2nu}
\end{equation}
where the explicit representation
\begin{equation}
 \Pi^{\mu2}_l= im^2\frac{\alpha}{\pi}\frac{1}{\cos\theta}\Bigl[\Lambda_1^\mu\,(G_1^l\sin\varphi+G_3^l\cos\varphi)+\Lambda_2^\mu\,(G_2^l\cos\varphi+G_4^l\sin\varphi)\Bigr]
\end{equation}
makes use of \Eqref{eq:lambda12}.  Using \Eqref{eq:Pi2nu} in
\Eqref{eq:M1}, the residual integration over $q_{\rm x}$ can be
performed and we obtain
\begin{equation}
 {\cal M}^{(p)}(k)=(2\pi)^3 \delta(k_{\rm z})\sum_{l\in\mathbb{Z}}\delta(k^0-2l\omega)\delta(k_{\rm y}-2l\omega\sin\theta)\,\frac{\epsilon_\mu^{*(p)}(k)}{\sqrt{2k^0}}
\Pi^{\mu2}_l(k,\tilde k)A(\tilde k_{\rm x})\,, \label{eq:M2}
\end{equation}
with $\tilde k^\mu\equiv(0,(k_{\rm x}-2l\omega\cos\theta)\vec{e}_{\rm x})$.

Taking into account the fact that the outgoing photon has positive
energy ($k^0>0$) and propagates on the light cone ($k_\mu k^\mu=0$),
and also because of the $\delta$ functions for the ${\rm y}$ and ${\rm z}$
momentum components, we identify $k^0\equiv2l\omega$ and
rewrite the $\delta$ function implementing energy conservation in
\Eqref{eq:M2} as follows,
\begin{equation}
 \delta(k^0-2l\omega)\ \to\ \delta_{l0}\,\delta(k_{\rm x})+ \Theta(l+0^+)\,\frac{1}{\cos\theta}\Bigl[\delta(k_{\rm x}-2l\omega\cos\theta)+\delta(k_{\rm x}+2l\omega\cos\theta)\Bigr]\,, \label{eq:delta}
\end{equation}
where $\Theta(.)$ is the Heaviside function.
Correspondingly, we have
\begin{multline}
 {\cal M}^{(p)}(k)=(2\pi)^3 \delta(k_{\rm z})\sum_{l=1}^{\infty}\frac{1}{\cos\theta}\Bigl[\delta(k_{\rm x}-2l\omega\cos\theta)+\delta(k_{\rm x}+2l\omega\cos\theta)\Bigr] \\ \times\delta(k_{\rm y}-2l\omega\sin\theta)\,\frac{\epsilon_\mu^{*(p)}(k)}{\sqrt{4l\omega}}\,\Pi^{\mu2}_l(k,\tilde k)A(\tilde k_{\rm x})\,, \label{eq:M3}
\end{multline}
with $k^\mu=(2l\omega,k_{\rm x},k_{\rm y},0)$, where we have
made use of the fact that the $l=0$ contribution vanishes: it scales
$\sim\delta(\vec{k})\,\frac{\Pi^{\mu2}_l(k,\tilde
  k)}{\sqrt{4l\omega}}\sim\delta(\vec{k})\,l^{3/2}\to0$ [cf. also
  \Eqref{eq:expansionparameters->} below].

When adapted to the particular kinematics in \Eqref{eq:M3} (cf. the arguments of the photon polarization tensor), the dimensionless parameters $\frac{k_1k_2}{4m^2}$, $\lambda$ and $\xi^2\lambda^2$
governing the expansion of the photon polarization tensor performed in Sec.~\ref{subsec:Pi} all vanish for the contribution $\sim\delta(k_{\rm x}-2l\omega\cos\theta)$.
For the contribution $\sim\delta(k_{\rm x}+2l\omega\cos\theta)$ they are non-zero and read
\begin{align}
 \frac{k_1k_2}{4m^2}\equiv \frac{k \tilde{k}}{4m^2} &\quad\to\quad\frac{1}{2}\left(\frac{2l\omega \cos\theta}{m}\right)^2, \nonumber\\
 \lambda&\quad\to\quad\left(\frac{2l\omega \cos\theta}{m}\right)\frac{\omega\cos\theta}{m}, \nonumber\\
 \xi^2\lambda^2&\quad\to\quad\left(\frac{e\mathfrak{E}}{m^2}\right)^2\left(\frac{2l\omega \cos\theta}{m}\right)^2\cos^2\theta. \label{eq:expansionparameters->}
\end{align}
Neglecting higher-order contributions of ${\cal O}(\frac{k_1k_2}{4m^2})\sim{\cal O}(\lambda)\sim{\cal O}(\frac{\omega^2}{m^2})$, our result will of course be fully governed by the remaining parameters $\xi_1^2\lambda^2$, $\xi_2^2\lambda^2$ and $\xi_1\xi_2\lambda^2$.

As a result, the number of merged photons with four wave-vector $k_f^\mu$
and polarization $p$ according to Fermi's golden rule is given by 
\begin{equation}
 {\cal N}^{(p)}(k_f)=\int\frac{d^3k}{(2\pi)^3}\,\bigl|{\cal M}^{(p)}(k)\bigr|^2= TL_{\rm y}L_{\rm z}\sum_{l=1}^{\infty}\frac{\bigl|\epsilon_\mu^{*(p)}(k_f)\Pi^{\mu2}_l(k_f,\tilde k_f)A(\tilde k_{f,{\rm x}})\bigr|^2}{4l\omega\cos\theta}\,, \label{eq:M4}
\end{equation}
with $k^\mu_f=2l\omega(1,-\cos\theta,\sin\theta,0)$, i.e., the
outgoing photon of energy $2l\omega$ propagates in
$(-\cos\theta,\sin\theta,0)$ direction.  Moreover, $\tilde
k^\mu_f=-4l\omega\cos\theta(0,\vec{e}_{\rm x})$ encodes the momentum
transfer from the field inhomogeneity, $T$ is the interaction time and
$L_{\rm y}L_{\rm z}$ is the interaction area transverse to the
inhomogeneity.  The total number of merged photons is
\begin{equation}
 {\cal N}(k_f)=\sum_{p}{\cal N}^{(p)}(k_f).  \label{eq:N}
\end{equation}
Obviously the dominant contribution is due to the merging of just two
laser photons, $l=1$, as higher photon processes are suppressed
by at least a factor of $\xi^2\lambda^2$.  Correspondingly,
\Eqref{eq:M3} can be written as
\begin{equation}
 {\cal N}^{(p)}(k_f)= TL_{\rm y}L_{\rm z}\frac{\bigl|\epsilon_\mu^{*(p)}(k_f)\Pi^{\mu2}_1(k_f,\tilde k_f)A(\tilde k_{f,{\rm x}})\bigr|^2}{4\omega\cos\theta}\bigl(1+{\cal O}(\tfrac{e^2{\mathfrak E}^2}{m^4}\tfrac{\omega^2}{m^2})\bigr)\,. \label{eq:M4a}
\end{equation}
We emphasize that the terms written out explicitly in \Eqref{eq:M4a} account for
the entire two-photon merging process.  We approximate the infinite
sum in \Eqref{eq:M4} by its contribution for $l=1$, and thereby
neglect merging processes of $2l$ laser photons with $l>1$.

Employing the substitutions $\varphi\to\varphi'$ and $\theta\to\pi-\theta$ in \Eqref{eq:a_12}, we introduce the following two vectors
\begin{align}
 \epsilon^{(1)\mu}(k_f)&=(0,-\sin\theta\cos\varphi',-\cos\theta\cos\varphi',-\sin\varphi'), \nonumber\\
 \epsilon^{(2)\mu}(k_f)&=(0,-\sin\theta\sin\varphi',-\cos\theta\sin\varphi',\cos\varphi'), \label{eq:epsilons}
\end{align}
with $\varphi'\in[0\ldots2\pi)$ fixed, to span the subspace transverse to the wave-vector $\vec{k}_f$ of the merged photon.
The two polarization degrees of freedom of the outgoing photon are then conveniently expressed in terms of the vectors $\epsilon^{(p)\mu}(k_f)$, with $p\in\{1,2\}$, representing linear polarization states in the particular basis characterized by a particular choice of $\varphi'$.
Polarizations other than linear can be obtained through linear combinations of the vectors~\eqref{eq:epsilons}.

We are now in a position to provide the explicit expressions of
the polarization tensor in \Eqref{eq:M4} contracted with a given
polarization vector of the outgoing photon, which read
\begin{multline}
 \epsilon_\mu^{*(1)}(k_f)\Pi^{\mu2}_l
= im^2\frac{\alpha}{\pi}\frac{1}{2\cos\theta}\Bigl[\sin\varphi'\,(G_1^l-G_2^l)-\sin(\varphi'+2\varphi)(G_1^l+G_2^l)  \\
-\cos(\varphi'+2\varphi)\,(G_3^l-G_4^l) -\cos\varphi'\,(G_3^l+G_4^l)\Bigr], \label{eq:epsilonPi}
\end{multline}
and
\begin{equation}
 \epsilon_\mu^{*(2)}(k_f)\Pi^{\mu2}_l=\epsilon_\mu^{*(1)}(k_f)\Pi^{\mu2}_l\big|_{\varphi'\,\to\,\varphi'-\frac{\pi}{2}}\,. \label{eq:epsilonPi2}
\end{equation}
Introducing the dimensionless field strengths $\varepsilon_i\equiv\frac{e\mathfrak{E}_i}{m^2}$ with $i\in\{1,2\}$, 
in particular the $l=1$ contribution to \Eqref{eq:epsilonPi} can be written as
\begin{multline}
 \epsilon_\mu^{*(1)}(k_f)\Pi^{\mu2}_{1}
=i(\omega \cos\theta)^2\frac{\alpha}{\pi}\frac{2}{15}\cos\theta\Bigl[-2i\sin(\varphi'+2\varphi)\,\varepsilon_1\varepsilon_2  \\
+\cos(\varphi'+2\varphi)\,(\varepsilon_1^2+\varepsilon_2^2) -\frac{11}{3}\cos\varphi'\,(\varepsilon_1^2-\varepsilon_2^2)\Bigr]\Bigl(1+{\cal O}(\tfrac{\omega^2}{m^2})\Bigr), \label{eq:epsilonPi_l=1}
\end{multline}
where we have made use of Eqs.~\eqref{eq:G_j^1} and
\eqref{eq:expansionparameters->}.

If $A(\tilde k_{f,{\rm x}})$ is either purely real or imaginary
valued, which is true for the field inhomogeneities symmetric in
  $\rm x$ to be considered below, the modulus squared can be split
and \Eqref{eq:M4a} be represented as follows,
\begin{equation}
 {\cal N}^{(p)}(k_f)= TL_{\rm y}L_{\rm
   z}\frac{\bigl|\epsilon_\mu^{*(p)}(k_f)\Pi^{\mu2}_1(k_f,\tilde
   k_f)|^2\,|A(\tilde k_{f,{\rm x}})\bigr|^2}{4\omega\cos\theta}\bigl(1+{\cal 
   O}(\tfrac{e^2{\mathfrak
     E}^2}{m^4}\tfrac{\omega^2}{m^2})\bigr)\,. \label{eq:M4b}
\end{equation}

The modulus squared of \Eqref{eq:epsilonPi_l=1} is obtained straightforwardly and reads
\begin{multline}
 \bigl|\epsilon_\mu^{*(1)}(k_f)\Pi^{\mu2}_1\bigr|^2
=(\omega \cos\theta)^4\frac{\alpha^2}{\pi^2}\frac{4}{225}\cos^2\theta\Bigl\{4(\varepsilon_1\varepsilon_2)^2
-\frac{22}{3}\cos\varphi'\cos(\varphi'+2\varphi)\,(\varepsilon_1^4-\varepsilon_2^4) \\
+\Bigl[\frac{121}{9}\cos^2\varphi'+\cos^2(\varphi'+2\varphi)\Bigr](\varepsilon_1^2-\varepsilon_2^2)^2\Bigr\}\Bigl(1+{\cal O}(\tfrac{\omega^2}{m^2})\Bigr),
\label{eq:modpisquared_1}
\end{multline}
while the analogous expression for the other polarization mode ($p=2$) follows from \Eqref{eq:epsilonPi2}.

Aiming at the total number of merged photons in the polarization basis characterized by a particular choice of $\varphi'$, we have to add the moduli squared corresponding to the two different polarization states [cf. Eqs.~\eqref{eq:N} and \eqref{eq:M4b}].
This results in
\begin{multline}
 \sum_{p=1}^2\bigl|\epsilon_\mu^{*(p)}(k_f)\Pi^{\mu2}_1\bigr|^2
=(\omega \cos\theta)^4\frac{\alpha^2}{\pi^2}\frac{8}{225}\cos^2\theta \\
\times\Bigl[4(\varepsilon_1\varepsilon_2)^2
-\frac{11}{3}\cos(2\varphi)\,(\varepsilon_1^4-\varepsilon_2^4)
+\frac{65}{9}(\varepsilon_1^2-\varepsilon_2^2)^2\Bigr]\Bigl(1+{\cal O}(\tfrac{\omega^2}{m^2})\Bigr),
\label{eq:modpisquared}
\end{multline}
which is completely independent of the choice of $\varphi'$, as it should.
Noteworthily, in case of circularly polarized incident
laser photons for which $\xi_1=\xi_2$ and thus
$\varepsilon_1=\varepsilon_2$, the contributions for both polarization
modes individually become independent of $\varphi$ and $\varphi'$;
cf. \Eqref{eq:modpisquared_1}. Equation \eqref{eq:modpisquared}
upon insertion into \Eqref{eq:N} and accounting for the prefactors
displayed in \Eqref{eq:M4a} represent a central result of this work.

\begin{figure}[h]
\center
\includegraphics[width=0.67\textwidth]{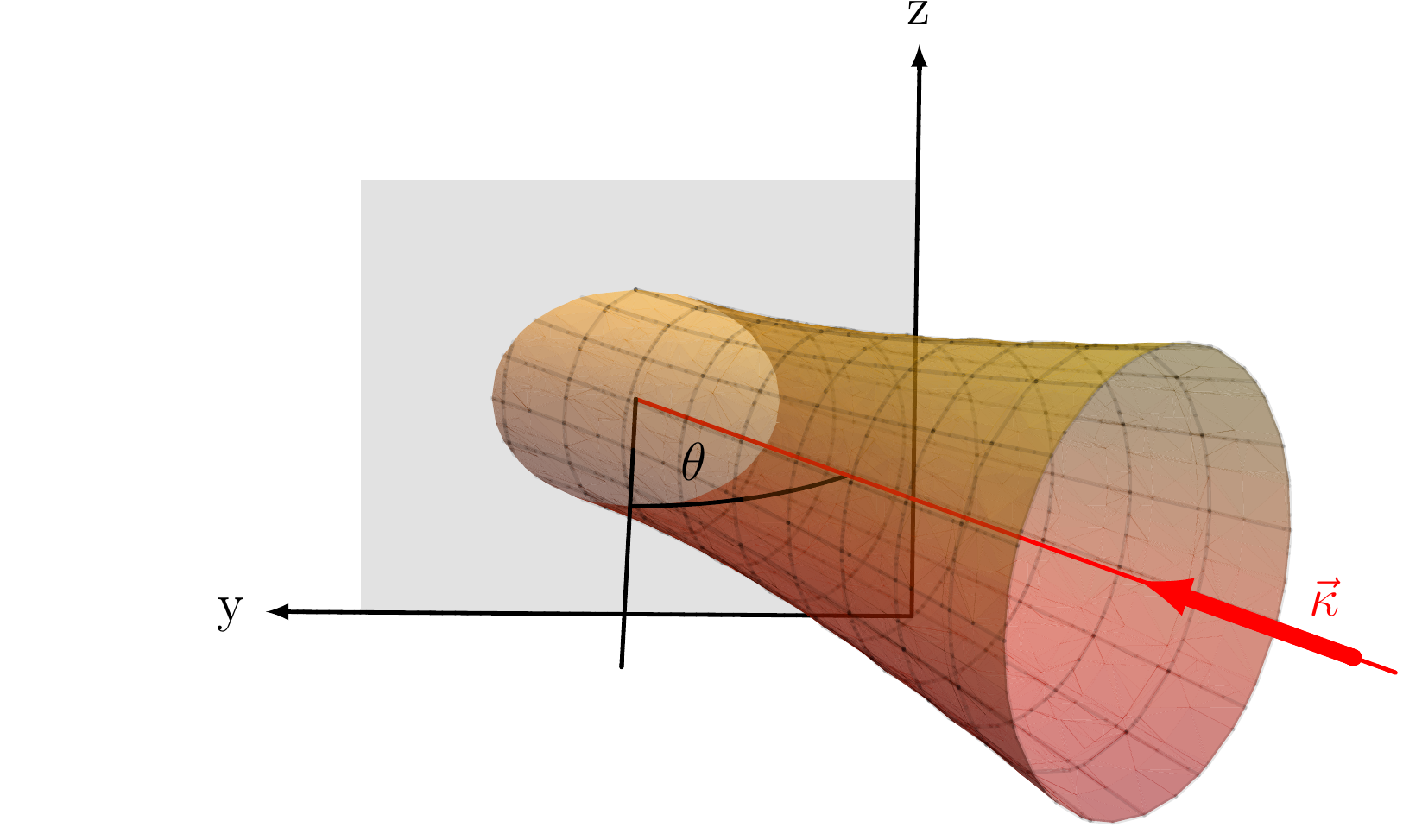} 
\caption{Sketch of the envelope of a Gaussian beam intersecting the ${\rm y}$--${\rm z}$ plane in
 the vicinity of its waist under an angle of $\theta$ (cf. also
 Fig.~\ref{fig:perspective}).  Given that the transverse cross-section
 area of the Gaussian beam at the beam waist is a circle of area
 $\sigma$, the intersection area is an ellipse with area $\frac{\sigma}{\cos\theta}$.}
\label{fig:yzSchnitt}
\end{figure}

Subsequently, we assume the probe laser to deliver incident laser
pulses of duration $\tau$, entering under an angle $\theta$ and
featuring a circular transverse beam profile.  The longitudinal
evolution of the probe laser pulses follows the envelope of a Gaussian
beam, with beam waist right at the intersection with the field
inhomogeneity.  We denote the transverse cross-section area at the
beam waist by $\sigma$.  Correspondingly, the transversal area $L_{\rm
  y}L_{\rm z}$ can be identified with the intersection area of such a
beam profile with the ${\rm y}$--${\rm z}$ plane, i.e., $L_{\rm y}L_{\rm
  z}=\frac{\sigma}{\cos\theta}$ (cf. Fig.~\ref{fig:yzSchnitt}).
Assuming that the magnetic field inhomogeneity is long-lived as compared to the
pulse duration $\tau$ of the probe laser, it is reasonable to consider
$\tau$ as a measure of the interaction time $T$, and set $T=\tau$.
Hence, we can make use of the following substitution,
\begin{equation}
 TL_{\rm y}L_{\rm z} \quad \to \quad \frac{\sigma\tau}{\cos\theta}\,. \label{eq:subst}
\end{equation}

\section{Results and Discussion} \label{seq:Ex+Res}

Let us now consider explicit examples of localized magnetic field inhomogeneities which can be tackled analytically.
We limit ourselves to two elementary shapes, characterized by just two parameters, namely an amplitude $B$ and a typical extension $w$.
For a Lorentz profile characterized by its full width at half maximum (FWHM),
\begin{equation}
 B({\rm x})=\frac{B}{1+(\frac{2{\rm x}}{w})^2}, \label{eq:Lorentz}
\end{equation}
the associated gauge field in position space can be determined by \Eqref{eq:Ax}. We obtain
\begin{equation}
 A({\rm x})=\frac{Bw}{2}\arctan\!\left(\frac{2{\rm x}}{w}\right),
\end{equation}
and Fourier transforming to momentum space via \Eqref{eq:Ak},
\begin{equation}
 A(q_{\rm x})=-i\frac{\pi Bw}{2q_{\rm x}}\,{\rm e}^{-\frac{|q_{\rm x}|w}{2}}. \label{eq:inh1}
\end{equation}

Analogously, for a Gaussian type inhomogeneity characterized by its full width at $1/{\rm e}$ of its maximum,
\begin{equation}
 B({\rm x})=B\,{\rm e}^{-\left(\frac{2{\rm x}}{w}\right)^2}, \label{eq:Gauss}
\end{equation}
we obtain
\begin{equation}
 A({\rm x})=\frac{\sqrt{\pi}Bw}{4}\,{\rm erf}\!\left(\frac{2{\rm x}}{w}\right),
\end{equation}
where ${\rm erf}(.)$ denotes the error function, and finally
\begin{equation}
 A(q_{\rm x})=-i\frac{\sqrt{\pi} Bw}{2q_{\rm x}}\,{\rm e}^{-\left(\frac{q_{\rm x}w}{4}\right)^2}. \label{eq:inh2}
\end{equation}
Equations~\eqref{eq:inh1} and \eqref{eq:inh2} share an overall prefactor $\sim(-i\frac{\sqrt{\pi} Bw}{2q_{\rm x}})$, but differ in the exponential decay.
For the Lorentz profile the decay is linear in $|q_{\rm x}|w$, while for the Gaussian inhomogeneity it is quadratic in this dimensionless parameter.

It is now straightforward to derive the number of merged laser photons, \Eqref{eq:M4b}, for these inhomogeneities.
The number of outgoing merged laser photons with polarization
$p=1$ and energy $2\omega$ reads [cf. Eqs.~\eqref{eq:M4b} and
  \eqref{eq:subst}]
\begin{multline}
 {\cal N}^{(1)}(k_f)
= w\sigma\tau(eB)^2(\omega w)\frac{\alpha\cos^2\theta}{57600\,\pi}\left\{\begin{array}{c}
            {\rm e}^{-(4\omega\cos\theta) w} \\
            \frac{1}{\pi}\,{\rm e}^{-\frac{1}{8}(4\omega\cos\theta)^2w^2}
        \end{array}\right\} \\
        \times\Bigl\{4(\varepsilon_1\varepsilon_2)^2
-\frac{22}{3}\cos\varphi'\cos(\varphi'+2\varphi)\,(\varepsilon_1^4-\varepsilon_2^4) \\
+\Bigl[\frac{121}{9}\cos^2\varphi'+\cos^2(\varphi'+2\varphi)\Bigr](\varepsilon_1^2-\varepsilon_2^2)^2\Bigr\}\Bigl(1+{\cal O}(\tfrac{\omega^2}{m^2})\Bigr),
 \label{eq:Np1}
\end{multline}
where the upper line in braces is the result for the Lorentz~\eqref{eq:Lorentz}
and the lower line that for the Gaussian~\eqref{eq:Gauss} profile, and
\begin{equation}
 {\cal N}^{(2)}(k_f)={\cal N}^{(1)}(k_f)\big|_{\varphi'\,\to\,\varphi'-\frac{\pi}{2}}\,. \label{eq:Np2}
\end{equation}
As the results for the Lorentz and Gaussian inhomogeneities -- apart
from the different exponential behavior -- are of very similar
structure, we find it convenient to adopt the two-component notation
employed in \Eqref{eq:Np1} in the remainder of this paper.
Equation \eqref{eq:Np1} exhibits several characteristic
dependencies on the involved parameters: as to be expected from the
underlying Feynman diagram, the leading order effect is proportional
to the square of the plane wave intensity, i.e., $\sim
\mathfrak{E}^4$, and to the square of the magnetic field $\sim B^2$.
In particular the latter dependence represents a comparatively
strong increase of the effect with an enhancement of the peak
magnetic background field. Other typical nonlinear phenomena such
as photon scattering off a magnetic field $\sim B^4$ or photon splitting
$\sim B^6$ are more strongly suppressed since the $B$ field scale is
measured in terms of the electron mass scale. On the other hand, the
inhomogeneous field has to provide the necessary momentum transfer
$\sim 4\omega\cos\theta$, and the effect is exponentially damped with $\sim(4\omega\cos\theta)w$.

In this respect, it is instructive to compare these expressions
with the number of photons experiencing quantum reflection
\cite{Gies:2013yxa} for the very same conditions, i.e., for incident
photons of the same energy, angle of incidence and polarization, and
exactly the same field inhomogeneities as in Eqs.~\eqref{eq:Lorentz}
and \eqref{eq:Gauss}.

For completeness, we note that in Ref.~\cite{Gies:2013yxa}, the field inhomogeneity was not accounted for exactly in the sense that the photon polarization tensor was evaluated {\it a priori} in the presence of the magnetic field inhomogeneity,
but rather the inhomogeneity was built in {\it a posteriori} by resorting to the result for a constant magnetic background field and using the constant-field expressions locally.
As argued in detail in~\cite{Gies:2013yxa}, such an approach is
justifiable for inhomogeneities whose typical scale of variation $w$
is much larger than the Compton wavelength $\lambda_c$ of the charged
virtual particles, i.e., $w\gg \lambda_c$.  Particularly in quantum
electrodynamics (QED), where the virtual particles are electrons,
$\lambda_c\approx2\cdot10^{-6}{\rm eV}^{-1}\approx3.9\cdot10^{-13}{\rm
  m}$, many field inhomogeneities available in the laboratory can be
dealt with along these lines.

Reference~\cite{Gies:2013yxa} identifies two situations for which the
calculations become particularly simple, corresponding to
special alignments of the incident photons' wave vector $\vec{k}$ and
polarization plane, the magnetic field $\vec{B}$, and the direction of
the inhomogeneity $\vec{\nabla}B$.
The one reconcilable with incident photons of four wave-vector
$\kappa^\mu=\omega(1,\cos\theta,\sin\theta,0)$ and
$\vec{B}\sim\vec{e}_{\rm z}$ is that with polarization vector in the
plane spanned by $\vec{\kappa}$ and $\vec{B}$, labeled by $\parallel$
in~\cite{Gies:2013yxa}.  To bring the $\parallel$ case
of~\cite{Gies:2013yxa} and the merging scenario discussed here into
full kinematic agreement, we specialize the quantum reflection formulae to
$\varangle(\vec{\kappa},\vec{B})=\frac{\pi}{2}$ and set
$\varphi=\varphi'=0$, $\varepsilon_1=0$ and
$\varepsilon_2=\frac{e\mathfrak{E}}{m^2}$ in Eqs.~\eqref{eq:Np1} and
\eqref{eq:Np2}, i.e., we specialize to incident laser photons
polarized linearly along ${\rm z}$, and look for induced outgoing
photons in the same polarization basis.  Incidentally, it can be shown
straightforwardly that the polarization direction is conserved under
these circumstances for quantum reflection (cf. \cite{Gies:2013yxa}),
i.e., the quantum reflected photons are still polarized along ${\rm z}$,
while for laser photon merging the induced outgoing photons are
polarized differently, namely their polarization vector lies in the
${\rm x}$--${\rm y}$ plane [cf. \Eqref{eq:Np-linpol1} below].

The number of quantum reflected photons ${\cal N}_{\rm Qref}$ is
obtained by multiplying the number of incident probe photons $N_{\rm
  probe}$ with the adequate reflection coefficient, given in Eqs.~(27)
and (29) of \cite{Gies:2013yxa}.
In order to allow for a more direct comparison with the merging result, we first rewrite $N_{\rm probe}$: The number of incident photons per pulse amounts to the ratio of the pulse energy of the probe laser $\cal E$ and its frequency $\omega$, i.e., $N_{\rm probe}=\frac{\cal E}{\omega}$.
The intensity $I_{\rm probe}$ at the focal spot, which is related to the electric field strength in the focus via $I_{\rm probe}={\mathfrak E}^2$, is determined by $I_{\rm probe}=\frac{\cal E}{\sigma\tau}$.
Hence, the number of probe photons can be expressed as $N_{\rm probe}=\frac{{\mathfrak E}^2\sigma\tau}{\omega}$, and -- neglecting corrections of ${\cal O}\bigl((\tfrac{eB}{m^2})^6\bigr)$ -- we finally obtain
\begin{equation}
{\cal N}_{\rm Qref}
= w\sigma\tau \frac{49\,\alpha}{129600\pi}
\biggl(\frac{eB}{m^2}\biggr)^4(e{\mathfrak E})^2(\omega w)\frac{1}{\cos^2\theta}
\left\{\begin{array}{c}
        \frac{1}{4}(1+\omega w\cos\theta)^2\,{\rm e}^{-2\omega w\cos\theta}\\
        \frac{1}{2\pi}\,{\rm e}^{-\frac{1}{4}(\omega w\cos\theta)^2} 
       \end{array}\right\} . \label{eq:Np-Qref}
\end{equation}
For the merging process, Eqs.~\eqref{eq:Np1} and \eqref{eq:Np2}, the same choice of parameters results in
\begin{equation}
{\mathcal N}^{(1)}
= w\sigma\tau\frac{49\,\alpha}{129600\,\pi}\biggl(\frac{e\mathfrak{E}}{m^2}\biggr)^4(eB)^2(\omega w)  \cos^2\theta \left\{\begin{array}{c}
            {\rm e}^{-4\omega w\cos\theta} \\
            \frac{1}{\pi}\,{\rm e}^{-2(\omega w\cos\theta)^2}
        \end{array}\right\}
 \Bigl(1+{\cal O}(\tfrac{\omega^2}{m^2})\Bigr) , \label{eq:Np-linpol1}
\end{equation}
while ${\mathcal N}^{(2)}=0$, such that ${\mathcal N}_{\rm
  merg}\equiv{\mathcal N}^{(1)}$. Both results exhibit an exponential
suppression with exponent $\sim w\omega\cos\theta=w\kappa_{\rm
  x}$, with $\kappa_{\rm x}$ being the momentum component of the
incident probe photons in the direction of the inhomogeneity
[cf. above \Eqref{eq:a_12}]. The suppression is more pronounced for the merging
process.  This can also be understood intuitively by recalling that
the momentum transfer from the inhomogeneity is $|2\kappa_{\rm x}|$
for the process of quantum reflection (cf. \cite{Gies:2013yxa}), while
it is twice as large, namely $|4\kappa_{\rm x}|$, for the merging of
two laser photons.

Another important point to notice is that in \Eqref{eq:Np-Qref} the
transition to large incidence angles $\theta\lesssim\pi/2$ provides a
convenient handle to damp the exponential suppression while at the
same time increasing the overall prefactor, which scales inversely
with $\cos^2\theta$.  Conversely, in \Eqref{eq:Np-linpol1} an
analogous increase of the angle of incidence to $\theta\lesssim\pi/2$
diminishes the overall prefactor $\sim\cos^2\theta$.
The ratio of Eqs.~\eqref{eq:Np-linpol1} and \eqref{eq:Np-Qref} can be derived straightforwardly, and reads
\begin{equation}
\frac{\mathcal{N}_{\rm merg}}{{\cal N}_{\rm Qref}}
\approx 4\,
\biggl(\frac{\mathfrak{E}}{B}\,\cos^2\theta\biggr)^2 
\left\{\begin{array}{c}
        \frac{1}{(1+\omega w\cos\theta)^{2}}\,{\rm e}^{-2\omega w\cos\theta} \\
        \frac{1}{2}\,{\rm e}^{-\frac{7}{4}(\omega w\cos\theta)^2}
       \end{array}\right\} . \label{eq:ratio}
\end{equation}
It is governed by just two dimensionless quantities, namely the product $\omega w \cos\theta$, measuring the width $w$ of the inhomogeneity in units of the inverse of the momentum component of the incident photons in $\vec{\nabla}B$ direction,
and $\mathfrak{E}/B\,\cos^2\theta$, i.e., the ratio of the field strength of the probe relative to that of the pump, augmented by an extra factor of $\cos^2\theta$.

It is now natural to ask for the conditions which have to be met such that photon merging dominates quantum reflection, i.e., $\mathcal{N}_{\rm merg}\geq{\cal N}_{\rm Qref}$.
Inserting this condition into \Eqref{eq:ratio}, we obtain
\begin{equation}
\frac{\mathfrak{E}}{B}\,\cos^2\theta
\geq \frac{1}{2}
\left\{\begin{array}{c}
        |1+\omega w\cos\theta|\,{\rm e}^{\omega w\cos\theta} \\
        \sqrt{2} \,{\rm e}^{\frac{7}{8}(\omega w\cos\theta)^2}
       \end{array}\right\}\geq \frac{1}{2}
\left\{\begin{array}{c}
        1 \\
        \sqrt{2}
       \end{array}\right\} , \label{eq:ratio2}
\end{equation}
where we made use of the fact that the expression on the right-hand
side of the first inequality is bounded from below by its value for
$\omega w\cos\theta=0$. 
The latter condition tells us that for the particular set-up
considered here, the yields for photon merging can dominate those for
quantum reflection only if the quantity $(\mathfrak{E}/B)\cos^2\theta$ is larger
than the numerical bounds given on the rightmost side of
\Eqref{eq:ratio2}.

\begin{figure}[h]
\center
\includegraphics[width=0.8\textwidth]{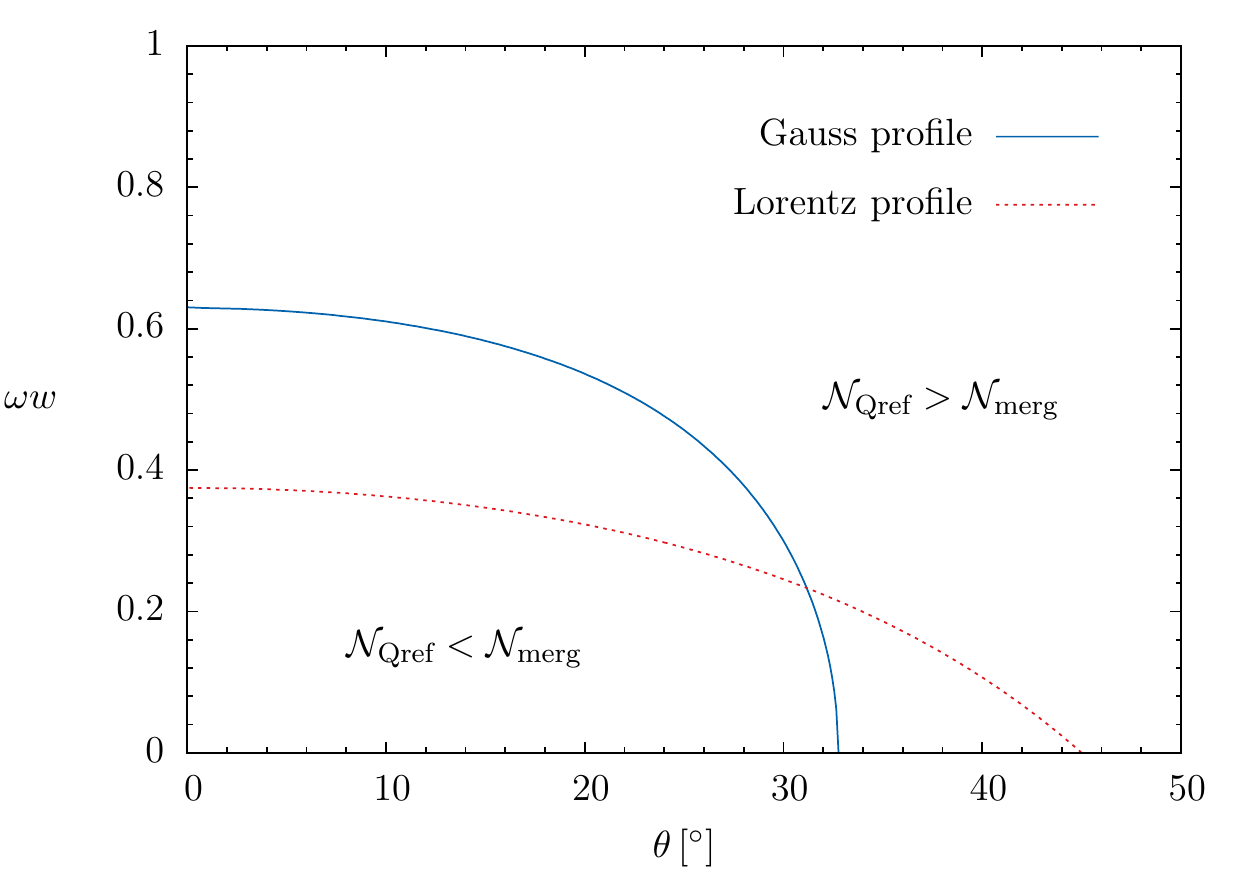} 
\caption{Choosing $\mathfrak{E}=B$ as an example,
we depict the regimes where photon merging dominates quantum
reflection and vice versa based on \Eqref{eq:ratio2}.  Photon
merging dominates quantum reflection in the regime in the lower left
bounded by the blue (solid) and red (dotted) lines for Gauss and
Lorentz type inhomogeneities, respectively.}
\label{fig:ratio}
\end{figure}

In Fig.~\ref{fig:ratio}, we exemplarily set $\mathfrak{E}=B$
which is a natural choice if all fields are provided by a
high-intensity laser system. We
investigate the implications of the first inequality in
\Eqref{eq:ratio2} as a function of $\theta$ and $\omega w$.
Obviously, for this choice of the field strengths laser photon merging
can only dominate quantum reflection if
$\cos\theta\geq\frac{1}{\sqrt{2}}$ $\leftrightarrow$
$\theta\leq45^\circ$ ($\cos\theta\geq 2^{-1/4}$ $\leftrightarrow$
$\theta\leq32.7^\circ$) for a Gaussian (Lorentzian) inhomogeneity.
Qualitatively speaking, the merging process tends to dominate for
small angles of incidence $\theta$ and small values of $\omega w$.
Equation~\eqref{eq:ratio} implies that this region (in the
$\theta$--$\omega w$ plane) can be enlarged by increasing the ratio of
$\mathfrak{E}/B$.

So far we only focused on the relative
importance of the two effects, but did not
provide absolute quantitative estimates.  Most obviously, as
both effects are suppressed by powers of $\frac{eB}{m^2}$ and
$\frac{e\mathfrak{E}}{m^2}$ [cf. Eqs.~\eqref{eq:Np-Qref} and
  \eqref{eq:Np-linpol1}], in order to increase them it is preferable
to enlarge the field strengths as much as possible.

Before providing some explicit quantitative estimates, let us briefly
discuss the generic features of \Eqref{eq:Np-linpol1} and confront it
with \Eqref{eq:Np-Qref}.
Consider the first derivative of the number of merged photons~\eqref{eq:Np-linpol1} with respect to $w\cos\theta$,
\begin{equation}
\frac{d{\mathcal N}_{\rm merg}}{d(w\cos\theta)}
\approx  \frac{2\,{\mathcal N}_{\rm merg}}{w\cos\theta}
\left\{\begin{array}{c}
            1-2\omega w\cos\theta \\
            1-2(\omega w\cos\theta)^2
        \end{array}\right\}\stackrel{!}{=}0 \quad \to\quad
 w\cos\theta = \frac{1}{2\omega}\left\{\begin{array}{c}
        1\\
        \sqrt{2}
       \end{array}\right\}.
 \label{eq:N_diff}
\end{equation}
Taking into account the sign of the second derivative, we find that 
the number of outgoing merged photons has a
maximum as a function of $w\cos\theta$ for the above values and reads 
\begin{equation}
{\mathcal N}_{\rm merg}\big|_{\rm max}
\approx \frac{\sigma\tau}{\omega}\frac{49\,\alpha}{129600\,\pi}\biggl(\frac{e\mathfrak{E}}{m^2}\biggr)^4(eB)^2\,\frac{1}{4}\left\{\begin{array}{c}
            {\rm e}^{-2} \\
            \frac{2}{\pi}\,{\rm e}^{-1}
        \end{array}\right\} . \label{eq:Nmax}
\end{equation}

Hence, keeping $w$ fixed, 
the number of merged photons increases monotonically as a
function of $\theta$ from its value for $\theta=0$ until it reaches
a maximum at $\theta = \arccos(\frac{1}{2\omega w})$ in case of the
Gaussian, and $\theta = \arccos(\frac{1}{\sqrt{2}\omega w})$ for the
Lorentz type inhomogeneity. Increasing $\theta$ even further, it
decreases rapidly until it reaches ${\mathcal N}_{\rm merg}=0$ at
$\theta=90^\circ$.

Conversely, for fixed $\omega$ the number of quantum reflected photons~\eqref{eq:Np-Qref} exhibits a monotonic increase throughout the interval from $\theta=0$ to $\theta=90^\circ$. 
Actually, ${\cal N}_{\rm Qref}$ even diverges for $\theta\to90^\circ$ due to the cosine squared term in its denominator, an unphysical feature which can be attributed to the unphysical limit of an infinitely long interaction of the probe photons and the inhomogeneity at ``grazing incidence'' $\theta\to90^\circ$.

Finally, we provide some rough estimates on the numbers of merged and
quantum reflected photons attainable in an all optical pump--probe
experiment based on high-intensity lasers.  Even though we have just
focused on a one-dimensional field inhomogeneity, as in
\cite{Gies:2013yxa} we exemplarily adopt the design parameters of the
two high-intensity laser systems to become available in Jena \cite{Jena}:
JETI~200 \cite{JETI200} ($\lambda=800{\rm nm}\approx4.06{\rm
  eV}^{-1}$, ${\cal E}=4{\rm J}\approx2.50\cdot10^{19}{\rm eV}$,
$\tau=20{\rm fs}\approx30.4{\rm eV}^{-1}$) as probe, and POLARIS
\cite{POLARIS} ($\lambda_{\rm pump}=1030{\rm nm}\approx5.22{\rm
  eV}^{-1}$, ${\cal E}_{\rm pump}=150{\rm
  J}\approx9.36\cdot10^{20}{\rm eV}$, $\tau_{\rm pump}=150{\rm
  fs}\approx228{\rm eV}^{-1}$) as pump.
This is meant to give a first order of magnitude estimate of the
number of induced outgoing photons.  Let us 
emphasize that it is certainly a rather crude approximation to adopt
the formula derived for a stationary, one-dimensional magnetic field
inhomogeneity of Gaussian type~\eqref{eq:Gauss} to mimic the field
inhomogeneity as generated in the focal spot of a high-intensity
laser.  Such an approximation ignores the {\it longitudinal modulation
  and evolution} of the pump laser pulse.  A more rigorous and refined
treatment in the context of an all optical pump--probe experiment
would require us to account also for the temporal structure and
evolution of field inhomogeneities. Fully accounting for pulse shape
dependencies has become a subject of increasing importance in
strong-field phenomenology with high-intensity lasers. Progress has
already been made, for instance, for the case of vacuum birefringence
\cite{DiPiazza:2006pr,Dinu:2014tsa}.

In generic high-intensity laser experiments the focal spot area cannot
be chosen at will, but is limited by diffraction.
Assuming Gaussian beams, the effective focus area is conventionally
defined to contain $86\%$ of the beam energy ($1/e^2$ criterion for
the intensity).  The minimum value of the beam diameter in the focus
is given by twice the laser wavelength multiplied with $f^\#$, the
so-called $f$-number, defined as the ratio of the focal length and the
diameter of the focusing aperture \cite{Siegman}; $f$-numbers as low
as $f^\#=1$ can be realized experimentally.  Thus, assuming both probe
and pump lasers to be focused down to the diffraction limit, the
attainable field strengths are of the order of
\begin{equation}
 \mathfrak{E}^2=I_{\rm probe}\approx \frac{0.86\,{\cal E}}{\tau\,\sigma}\,, \quad B^2=2I_{\rm pump}\approx2\,\frac{0.86\,{\cal E}_{\rm pump}}{\tau_{\rm pump}\,\sigma_{\rm pump}}\,,
\label{eq:EBpump}
\end{equation}
with $\sigma\approx\pi\lambda^2$ and $\sigma_{\rm pump}\approx\pi\lambda_{\rm pump}^2$. The additional factor of two in the definition of $B$ accounts for the fact that, focusing on a purely magnetic field inhomogeneity, the entire laser intensity is considered to be available in terms of a magnetic field, as could, e.g., be realized by superimposing two counter propagating laser beams.

In the most straightforward experimental setting to imagine, the pump laser beam propagates along the ${\rm y}$ axis, while its transversal profile,
parametrized by the coordinate ${\rm x}$, evolves along the well-defined envelope of a Gaussian beam, and in the vicinity of the beam waist is to be understood as constituting the Gaussian field inhomogeneity~\eqref{eq:Gauss} of width $w\approx2\lambda_{\rm pump}$.

For beams focused down to the diffraction limit, the Rayleigh length 
is given by the wavelength of the beam multiplied with a factor of
$\pi$ \cite{Siegman}, i.e., for the pump, $z_{\rm R}=\pi\lambda_{\rm
  pump}$.  Hence, over distances of the order of several wave lengths
$\lambda_{\rm pump}$ about the beam waist, the beam diameter remains
approximately constant along $\vec{e}_{\rm y}$ and an experimental
setting resembling Fig.~\ref{fig:perspective} is conceivable.

\begin{figure}[h]
\center
\includegraphics[width=0.8\textwidth]{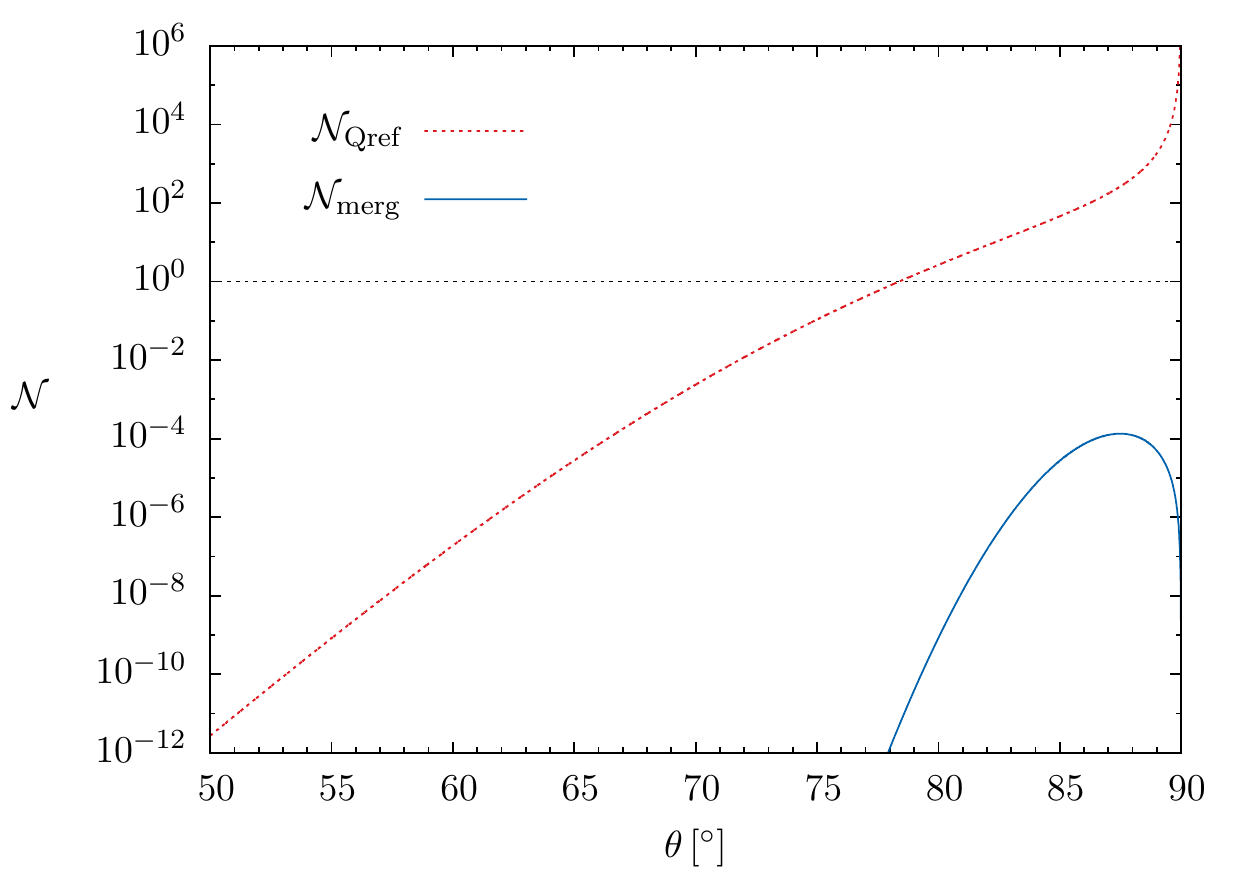} 
\caption{Number of induced outgoing photons per shot ${\cal N}_{\rm
    merg}$ due to the effects of laser photon merging and quantum
  reflection as a function of $\theta$, adopting the design parameters of the
  Jena high-intensity laser systems, JETI~200 and Polaris (cf. main
  text).  The horizontal dashed line shows where the number of induced
  outgoing photons per shot becomes one.  For quantum reflection this
  is the case for $\theta\geq78^\circ$ \cite{Gies:2013yxa}.
  Conversely, the number of outgoing merged photons reaches a maximum
  at $\theta\approx87^\circ$ and stays below ${\cal N}_{\rm
    merg}\big|_{\rm max}\approx1.3\cdot10^{-4}$ throughout the
  interval $0\leq\theta\leq90^\circ$; cf. \Eqref{eq:Nmax} for the
  Gaussian inhomogeneity and the discussion below.  For completeness,
  we note that ${\cal N}_{\rm
    Qref}\bigr|_{\theta=0}\approx3\cdot10^{-29}$ while ${\cal N}_{\rm
    merg}\bigr|_{\theta=0}\approx3\cdot10^{-228}$ .}
\label{fig:qualitative}
\end{figure}

In Fig.~\ref{fig:qualitative}, we plot the number of induced outgoing photons for both effects as a
function of $\theta$. The respective results are obtained
straightforwardly by plugging the design parameters of the Jena
high-intensity laser systems JETI~200 and Polaris given above into
\Eqref{eq:EBpump} and the lower components of Eqs. \eqref{eq:Np-Qref},
\eqref{eq:Np-linpol1} and \eqref{eq:Nmax}. 

Obviously, for this particular all-optical experimental setup the
photon merging process is substantially suppressed in comparison with
quantum reflection.  As detailed below \Eqref{eq:Np-linpol1}, the
differences observed in Fig.~\ref{fig:qualitative} can be attributed
to the different scaling of Eqs. \eqref{eq:Np-Qref} and
\eqref{eq:Np-linpol1} with $\cos^2\theta$.  While quantum reflection
receives an overall enhancement with $\sim\frac{1}{\cos^2\theta}$ for
large angles of incidence $\theta\lesssim90^\circ$, photon merging
becomes maximal if the condition~\eqref{eq:Nmax} is met (for the
JETI~200 -- Polaris setup this is the case for an angle of
$\theta\approx 87^\circ$, wherefore ${\cal N}_{\rm merg}\big|_{\rm
  max}\approx1.3\cdot10^{-3}$) and dies off to zero for
$\theta\to90^\circ$.

In practice, an all-optical setup designed to benefit from the
geometric noise reduction will work at a reflection angle near or
somewhat above $\theta \simeq 80^\circ$. For parameters similar to
the ones studied here, photon merging then is clearly a negligible
background to the quantum reflection signal. Nevertheless, because
of its different polarization and frequency dependence, appropriate
filtering techniques could still render photon merging detectable in the long run.

\section{Conclusions and Outlook} \label{seq:Con+Out}

In this paper we have studied laser photon merging in the presence of
a one dimensional, stationary magnetic field inhomogeneity.  We have
in particular confronted the number of outgoing merged photons with
the number of quantum reflected photons for the same conditions and
discussed in detail the similarities and differences of the two
effects.  Sticking to the design parameters of the high-intensity
laser facilities to be available in Jena, consisting of a petawatt and
a terawatt class laser system, we have provided a first rough estimate
of the number of merged photons to be potentially attainable in an
all-optical experiment. Our results confirm that the quantum
reflection signal is a most promising candidate for the discovery of
quantum vacuum nonlinearities under controlled laboratory conditions
with high-intensity lasers. In particular, it dominates photon
merging in a wide parameter range.

The expression for the photon merging number is determined most
straightforwardly from
the photon polarization tensor in a plane wave background.  Actually,
the main difficulty in determining the number of outgoing merged laser
photons is the problem of finding a convenient and controllable
expansion of the photon polarization tensor, allowing us to represent
our results in concise expressions.  This has led us to adopt a novel
expansion strategy to obtain analytical insights into the photon
polarization in plane wave backgrounds. We believe that this
representation will also be useful in many other strong field physics
questions beyond the merging process.

Of course, a natural extension of our present study in the future
would be the investigation of the photon merging process in more
generic, time-dependent inhomogeneities.  Such a study is necessary to
allow for definitive answers about the the numbers of outgoing merged
photons attainable in the focal spot of high-intensity lasers, taking
into account the full longitudinal evolution of the pump laser pulse.

\section*{Acknowledgments}

We are particularly indebted to Maria~Reuter for creating
Figs.~\ref{fig:perspective} and \ref{fig:yzSchnitt}.  FK is
grateful to Matt~Zepf for many interesting and enlightening discussions.
HG acknowledges support by the DFG under grants Gi 328/5-2 (Heisenberg
program) and SFB-TR18. RS acknowledges support by the Ministry of Education
and Science of the Republic of Kazakhstan.

\end{document}